\documentclass[journal=jctcce,manuscript=article]{achemso}  
\usepackage{amsfonts}
\usepackage{bm}
\usepackage{setspace}
\usepackage{graphicx}
\usepackage{color}
\usepackage{multirow}
\usepackage{verbatim}
\usepackage{float}
\usepackage{makecell}
\usepackage[version=3]{mhchem}
\usepackage{color}

\usepackage[ruled,linesnumbered]{algorithm2e}

\usepackage{newunicodechar}

\newunicodechar{≤}{\ensuremath{\leq}}
\newunicodechar{≥}{\ensuremath{\geq}}

\newcommand{\be}{\begin{equation}}
\newcommand{\ee}{\end{equation}}
\newcommand{\bea}{\begin{eqnarray}}
\newcommand{\eea}{\end{eqnarray}}
\newcommand{\bsube}{\begin{subequations}}
\newcommand{\esube}{\end{subequations}}

\makeatletter

\newcommand{\Rmnum}[1]{\expandafter\@slowromancap\romannumeral #1@}
\makeatother

\author{Yi-Fan Yao}
\author{Neil Qiang Su}
\email{nqsu@nankai.edu.cn}
\affiliation{Center for Theoretical and Computational Chemistry, Frontiers Science Center for New Organic Matter, State Key Laboratory of Advanced Chemical Power Sources, Key Laboratory of Advanced Energy Materials Chemistry (Ministry of Education), Department of Chemistry, Nankai University, Tianjin 300071, China}

\title{Enhancing Reduced Density Matrix Functional Theory Calculations by Coupling Orbital and Occupation Optimizations}

\begin{document}

\begin{tocentry}
\begin{center}
\includegraphics[width=8.0cm,height=4.5cm]{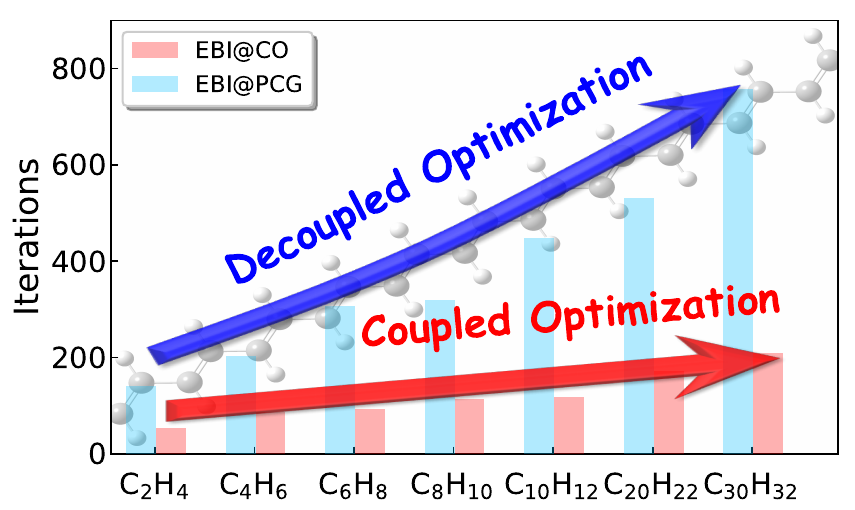}
\par\end{center}
\end{tocentry}

\begin{abstract}
Reduced density matrix functional theory (RDMFT) calculations are usually implemented in a decoupled manner, where the orbital and occupation optimizations are repeated alternately. Typically, orbital updates are performed using the unitary optimization method, while occupations are optimized through the explicit-by-implicit (EBI) method. The EBI method addresses explicit constraints by incorporating implicit functions, effectively transforming constrained optimization scenarios into unconstrained minimizations. Although the unitary and EBI methods individually achieve robust performance in optimizing orbitals and occupations, respectively, the decoupled optimization methods often suffer from slow convergence and require dozens of alternations between the orbital and occupation optimizations. To address this issue, this work proposes a coupled optimization method that combines unitary and EBI optimizations to update orbitals and occupations simultaneously at each step.  To achieve favorable convergence in coupled optimization using a simple first-order algorithm, an effective and efficient preconditioner and line search are further introduced. The superiority of the new method is demonstrated through numerous tests on different molecules, random initial guesses, different basis sets and different functionals. It outperforms all decoupled optimization methods in terms of convergence speed, convergence results and convergence stability. Even a large system like $\mathrm{C_{60}}$ can converge to $10^{-8}$ au in 154 iterations, which shows that the coupled optimization method can make RDMFT more practical and facilitate its wider application and further development. 
\end{abstract}

\maketitle

%
%

\section{Introduction}

Reduced Density Matrix Functional Theory (RDMFT) is a novel functional theory in quantum chemistry, introduced by Gilbert in 1975 \cite{Gilbert1975prb}. It has garnered significant interest over the years  \cite{Gilbert1975prb,Levy1979pnas,Muller1984rpa,
GU1998prl,Pernal2005prl,Gritsenko2005jcp,Rohr2008jcp,Sharma2008prb,Lathiotakis2009pra,Piris2010jcp,
Sharma2013prl,Schade2017,Christian2018,Schilling2019prl,polar2020jctc,Piris2021prl,Gibney2021jpcl,
Yao2021jpcl,Ai2023jcp,Gibney2022jctc,Gibney2023prl} for its potential to address the inherent limitations of the widely used density functional theory (DFT)\cite{HK64,KS65,PY1989,Cohen2008science,Xu2005jcp,Dreizler2012,Su2018pnas}. Taking the one-electron reduced density matrix (1-RDM) $\gamma_\sigma$ \cite{Lowdin1955}, as the basic variable, the total energy reads
\begin{equation}
    E[\gamma _\sigma] = T[\gamma_\sigma] + E_{ext}[\gamma_\sigma]+E_H[\gamma_\sigma]+E_{XC}[\gamma_\sigma], \label{energy 1}
\end{equation}
where $T\left[\gamma_\sigma\right]=-\frac{1}{2}\sum_\sigma\int \gamma_\sigma(\mathbf{r},\mathbf{r'})\nabla ^2\delta(\mathbf{r}-\mathbf{r'})d\mathbf{r}d\mathbf{r'}$, $E_{\mathrm{ext}}\left[\gamma_\sigma\right]=\int v_{ext}(\mathbf{r})\rho(\mathbf{r})d\mathbf{r}$, $E_H\left[\gamma_\sigma\right]=\frac{1}{2}\int \frac{\rho(\mathbf{r})\rho(\mathbf{r'})}{|\mathbf{r}-\mathbf{r'}|}d\mathbf{r}d\mathbf{r'}$ are the well-defined kinetic energy, (non-)local external energy and classical Coulomb energy, respectively, with $\rho(\mathbf{r})=\sum_\sigma \gamma_\sigma(\mathbf{r},\mathbf{r})$. $E_{\mathrm{XC}}\left[\gamma_\sigma\right]$ is the unknown exchange-correlation (XC) energy that needs to be approximated. The eigenvalues of 1-RDM, i.e. the orbital occupations, can be fractional to better capture strong correlation effects, avoiding the use of multireference wave functions that are computationally intractable. These advantages make RDMFT a promising research field.

Despite its theoretical advantages, RDMFT suffers from high computational cost and low convergence accuracy, which hinder its wide application in many fields \cite{Pernal2005prl,piris2009iterative,Schilling2019prl,Theophilou2016,Baldsiefen2013ctc,Yao2021jpcl}.  Given the spectral representation of $\gamma_\sigma$
\begin{equation}
    \gamma_\sigma = \sum_i^M |\psi  _i^\sigma\rangle n_i^\sigma \langle \psi _i^\sigma |,
\end{equation}
the ground-state energy should be obtianed through the minimization with respect to natural orbitals (NOs) ${\psi_p^\sigma}$ and  occupation numbers (ONs) ${n_p^\sigma}$ under the ensemble $N$-representability constraints \cite{Lowdin1955,Coleman1963rmp,Valone1980jcp}
\begin{equation}
\langle \psi_i^\sigma | \psi_j^\sigma \rangle =\delta_{ij},\label{orthogonal}
\end{equation}
\begin{equation}
\quad 0\leq n_i^\sigma \leq 1, \label{limit}
\end{equation}
\begin{equation}
 t^\sigma =  \sum_i^M n_i^\sigma - N_0^\sigma = 0, \label{N-represent}
\end{equation}
where $N_0^\sigma$ is  the electron number of $\sigma$ spin in the system. In this work, $K$ denotes the dimension of the basis set, while $M$ signifies the number of molecular orbitals, retained after removing linear dependencies via L{\"{o}}wdin orthonormalization \cite{lowdin1950jcp}. Several approaches have been developed to address the computational challenge of RDMFT from different perspectives \cite{cances2008projected, PhysRevA.14.36, RevModPhys.80.3, ferre2016density}, such as transforming the problem into an eigenproblem\cite{Pernal2005prl, giesbertz2010aufbau, PhysRevB.77.235121} and using the resolution-of-identity (RI) approximation\cite{lemke2022efficient}. Despite these, more efforts are still needed to make RDMFT a practical method. Compared with previous optimization methods that alternately optimize NOs and ONs, this work presents a novel method that couples NO and ON optimization. This new method not only enhances the convergence rate of RDMFT calculations, but also alleviates the problem of getting stuck in local minima. Using this method, we can achieve an energy convergence of 1e-8 within 200 steps for the $\mathrm{C_{60}}$ system, which greatly improves the computing power of RDMFT.

This article is organized as follows. In Section \ref{theory}, we review the energy functional in RDMFT and the traditional decoupled optimization methods. Then, we develop the coupled optimization method and present the algorithm. In Section \ref{comp}, we describe the computational details of our implementation. In Section \ref{results}, we compare the convergence speed and energies of the new optimization method with the decoupled methods. We also evaluate the performance of the new method on different systems, basis sets, and functionals. Finally, in Section \ref{conc}, we summarize our main findings and conclusions.

\section{Theory \label{theory}}

Throughout this paper, we adopt the standard notation for orbital index labels: the indices $p$, $q$, $a$,  $b$, $i$, $j$, $k$ and $l$  denote molecular orbitals, the indices $\mu $, $v$, $\kappa$  and $\theta $  denote atomic orbitals, and the index $\sigma$ denotes the spin of orbital, which can be either $\alpha$ or $\beta$.

\subsection{Energy functionals in RDMFT}

By inserting $\psi_i^\sigma = \sum_\mu^K C_{\mu i}^\sigma\phi _\mu^\sigma$, the energy functional of Eq. \ref{energy 1} can be expressed as
\begin{equation}
    \begin{split}
        &E[\{\psi^\sigma_p\}, \{n_p^\sigma\}] = \sum_\sigma^{\alpha, \beta}\sum_i^{M}\sum_{\mu  v}^K C_{ \mu i}^\sigma h_{\mu v}^\sigma C_{vi}^\sigma n_i^\sigma\\
        &+\frac{1}{2}\sum_{\sigma}^{\alpha, \beta}   \sum_{ij}^M\sum_{ \mu v \kappa  \theta}^K C_{\mu i}^\sigma C_{ vi}^\sigma (\mu v|\kappa\theta)C_{ \kappa j}^\sigma C_{\theta j}^\sigma n_i^\sigma n_j^\sigma+E_\mathrm{XC}.\\
    \end{split} \label{energy 2}
\end{equation}
where $\{\phi_\mu^\sigma\}$ and $C^\sigma$  are atomic orbitals and molecular orbital expansion coefficients, 
respectively. The three terms in Eq. \ref{energy 2} correspond to $T\left[\gamma_\sigma\right]+E_{\mathrm{ext}}\left[\gamma_\sigma\right]$,    $E_H\left[\gamma_\sigma\right]$ and $E_{XC}\left[\gamma_\sigma\right]$ in Eq. \ref{energy 1} respectively.
Here $h_{\mu v}^\sigma$ is the one-electron integral,
\begin{equation}
   h_{\mu v}^\sigma = \int {\phi^\sigma_\mu}^*(\mathbf{r})[-\frac{1}{2}\nabla ^2 +V_{ext}(\mathbf{r})]\phi^\sigma_v(\mathbf{r}) d\mathbf{r},
\end{equation}
$V_{ext}(r)$ is the external potential, $\nabla ^2$ is the Laplace operator. $(\mu v|\kappa \theta)$ is the two-electron integral
\begin{equation}
    (\mu v|\kappa \theta) = \iint \frac{{\phi_\mu^\sigma}^* (\mathbf{r})\phi_v^\sigma(\mathbf{r}) \phi_\kappa^\sigma(\mathbf{r'}) {\phi_\theta^\sigma}^*(\mathbf{r'})}{|\mathbf{r}-\mathbf{r'}|} d\mathbf{r}d\mathbf{r'}.
\end{equation}
The exchange-correlation (XC) term in the following derivation has the form
\begin{equation}
    E_\mathrm{XC} \!=\! -\frac{1}{2}\sum_{\sigma}^{\alpha, \beta} \sum_{ij}^{M}\sum_{ \mu v \kappa  \theta}^K C_{ \mu i}^\sigma C_{v j}^\sigma (\mu v|\kappa\theta)C_{ \kappa i}^\sigma C_{\theta j}^\sigma f(n_i^\sigma ,n_j^\sigma).
\end{equation}
This form is widely used in many functionals, such as the M\"{u}ller \cite{Muller1984rpa} and Power \cite{Sharma2008prb} functionals, which are utilized in this work. The Power functional  \cite{Sharma2008prb} is
\begin{equation}
  E_{\mathrm{XC}}^{\mathrm{Power}}[\gamma_\sigma] = -\frac{1}{2}\sum_{\sigma}^{\alpha,\beta}\iint \frac{|\gamma_\sigma^m (\mathbf{r}_1,\mathbf{r}_2)|^2 }{|\mathbf{r}_1-\mathbf{r}_2|}\mathrm{d}\mathbf{r}_1 \mathrm{d}\mathbf{r}_2,
\end{equation}
The $\omega$P22 functional \cite{Ai2022jpcl,Ai2023jcp} takes the form of
\begin{equation}
  E_{\mathrm{XC}}^{\mathrm{\omega P22}} =  E_{\mathrm{XClr}}\left[\gamma_\sigma\right]+ E_{\mathrm{Xsr}}^{\mathrm{KS}}\left[\rho_\sigma\right]+U_{\mathrm{Csr}}^{\mathrm{KS}}\left[\rho_\sigma\right],
\end{equation}
where 
\begin{equation}
  E_{\mathrm{XClr}}\left[\gamma_\sigma\right]=\frac{1}{2} \iint \frac{\operatorname{erf}\left(\omega\left|\mathbf{r}_1-\mathbf{r}_2\right|\right)}{\left|\mathbf{r}_1-\mathbf{r}_2\right|} \rho\left(\mathbf{r}_1\right) \rho_{\mathrm{XC}}^{\mathrm{RDM}}\left(\mathbf{r}_2 \mid \mathbf{r}_1\right) \mathrm{d} \mathbf{r}_1 \mathrm{~d} \mathbf{r}_2,
  \end{equation}
and $\rho_{\mathrm{XC}}^{\mathrm{RDM}}$  is
\begin{equation}
  \rho_{\mathrm{XC}}^{\text {RDM }}\left(\mathbf{r}_1, \mathbf{r}_2\right)=-\sum_\sigma \frac{\left|\gamma_\sigma^m\left(\mathbf{r}_1, \mathbf{r}_2\right)\right|^2}{\rho\left(\mathbf{r}_1\right)}.
  \end{equation}
The short-range B88 functional \cite{B88,Iikura2001jcp} and the LYPsr functional \cite{LYPr} are utilized for $E_{\rm{Xsr}}^{\rm{KS}}$ and $U_{\rm{Csr}}^{\rm{KS}}$ respectively. The power $m$ and the range-separation parameter $\omega$ in $\omega$P22 take the values of 0.6 and 0.45, respectivaly \cite{Ai2023jcp}. In addition to the aforementioned functionals, the optimization methods in this work can also handle other functionals.

\subsection{Decoupled optimization} 

In RDMFT, NO and ON optimizations are usually performed in the decoupled manner. That is, the NOs are optimized while keeping the ONs fixed, and vice versa. This process is repeated alternately until convergence.

The orthonormality of natural orbitals (NOs) in Eq. \ref{orthogonal} is preserved by applying a unitary transformation on $C^\sigma$ \cite{Traian2009}
\begin{equation}
C_{i+1}^\sigma =  C_i^\sigma e^{R^\sigma} =C_i^\sigma \sum_{k=0}^\infty \frac{(R^\sigma)^k}{k!}, \label{c mat}
\end{equation}
where $R^\sigma$ is an antisymmetric matrix. This allows the optimization of NOs to be implemented by updating $R^\sigma$ at each step.

ONs can be expressed as cosine functions to satisfy the constraints in Eq. \ref{limit}, i.e., $n_p^\sigma=\mathrm{cos}^2(x_p^\sigma)$, while the constraint of Eq. \ref{N-represent} on the sum of ONs can be handled by the Lagrange multiplier (LM) method or its variant, the augmented Lagrange multiplier method (ALM) \cite{bertsekas2014constrained}. The objective functions of LM and ALM take the forms:
\begin{equation}
\label{eq:LM}
\mathcal{L}=E\left[\gamma_\sigma\right]+ \sum_\sigma^{\alpha,\beta} \lambda^\sigma t^\sigma\!\left(\{n_p^\sigma\}\right),
\end{equation} 
\begin{equation}
\label{eq:ALM}
\mathcal{L}_a\!=\!E\left[\gamma_\sigma\right]\!+\! \sum_\sigma^{\alpha,\beta} \lambda_k^\sigma t^\sigma\!\left(\{n_p^\sigma\}\right)\!+\!\sum_\sigma^{\alpha,\beta} \frac{c_k^\sigma}{2} |t^\sigma\!\left(\{n_p^\sigma\}\right)|^2.
\end{equation}
In LM, $\lambda^\sigma$ is the Lagrange multiplier that is jointly optimized with $\{x_p^\sigma\}$,
while in ALM, $\{x_p^\sigma\}$ are optimized for a series of fixed $c_k^{\sigma}$ and $\lambda_k^{\sigma}$ that are updated by $c_k^{\sigma}=2^k$ and $\lambda_k^{\sigma}=\lambda_{k-1}^{\sigma}+c_{k-1}^{\sigma}t^{\sigma}\left(\{x_p^{\sigma}\}_{k-1}\right)$ \cite{bertsekas2014constrained}. However, as shown in our previous work, both methods have drawbacks in dealing with the constraints of ONs. LM requires a second-order method to converge, and its convergence stability depends heavily on the initial guess, while ALM needs many iterations to satisfy the constraints on the ONs \cite{nocedal2006numerical}.

The explicit-by-implicit (EBI) method \cite{Yao2021jpcl,Yao2022jpca} was proposed in Su group to address the convergence issues of LM/ALM. EBI parameterizes ONs with sigmoid functions without introducing redundant variables,
\begin{equation}
n_p^\sigma = s(x_p^\sigma, \mu^\sigma), \label{single implicit}
\end{equation}
where $\{x_p^\sigma\}$ are unconstrained variables, and $\mu^\sigma$ are implicit functions of $\{x_p^\sigma\}$. In this work, the error function is used to represent ONs as $n_p^\sigma = (\mathrm{erf}(x_p^\sigma+\mu^\sigma)+1)/2$. This approach has several advantages for constrained optimization. First, for any $\{x_p^\sigma\}$, the constraint of Eq. \ref{N-represent} is a monotonic function of $\mu^\sigma$, which can be easily solved to obtain the ONs that satisfy the constraint. This means that, unlike LM or ALM, which only satisfy the constraint at convergence, EBI satisfies the constraint at every step of the optimization, which also facilitates the coupled optimization of NOs and ONs introduced below. Second, EBI can achieve fast convergence using a first-order optimization method \cite{Yao2022jpca}, and the convergence results are stable and robust to the initial guess, which is essential for applying RDMFT to large systems. Besides the error function, other sigmoid functions could also be considered for parameterizing ONs, such as the Fermi-Dirac distribution \cite{freysoldt2009direct}, which shares similar advantages in this context.

\subsection{Coupled optimization} 

The decoupled optimization method often suffers from slow convergence, and it usually requires dozens of alternations between the NO and ON optimizations. To address this issue, in this work, we propose a coupled optimization method that combines unitary and EBI optimizations to update NOs and ONs simultaneously at each step. 

When the coupled method is carried out with first-order numerical optimization algorithms, the first derivatives of the energy functional with respect to both $R^\sigma$ and $\{x_p^\sigma\}$ are required. They are the first derivatives of $E$ with respect to $R^\sigma$
\begin{equation}
    \begin{split}
        \left.\frac{\partial E}{\partial R^\sigma_{pq}}\right|_{R^\sigma=0}
        =\sum_\mu^K \left(C^\sigma_{\mu p} \frac{\partial E}{\partial C^\sigma_{\mu q}}-\frac{\partial E}{\partial C^\sigma_{\mu p}}C^\sigma_{\mu q}\right), \label{M mat}
    \end{split}
\end{equation}
where
\begin{equation}
  \begin{split}
    &\frac{\partial E}{\partial C_{\mu i}^{\sigma}}= 2\sum_{   v}^K h_{\mu v}^\sigma C_{vi}^\sigma n_i^\sigma
        +2\sum_{j}^{M} \sum_{  v \kappa  \theta}^K C_{v i}^\sigma  (\mu v|\kappa\theta)C_{ \kappa j}^\sigma C_{\theta j}^\sigma n_i^\sigma n_j^\sigma\\
        &-2 \sum_{j}^{M}\sum_{  v \kappa  \theta}^K  C_{ vj}^\sigma (\mu v|\kappa\theta)C_{ \kappa i}^\sigma C_{\theta j}^\sigma f(n_i^\sigma ,n_j^\sigma).\\
  \end{split}
\end{equation}
And the first derivatives of $E$ with respect to $x_p^\sigma$ \cite{Yao2021jpcl,Yao2022jpca} are
\begin{equation}
    \label{eq:EBI1d}
    \frac{\partial E}{\partial x_p^\sigma} =\sum_{q=1}^{M} \frac{\partial E}{\partial n_q^\sigma}\frac{\partial n_q^\sigma}{\partial x_p^\sigma},
\end{equation}
where
\begin{subequations}
	\begin{align}
    \begin{split}
        \frac{\partial E}{\partial n_q^\sigma} = &\sum_{ \mu  v}^K C_{ \mu q}^\sigma h_{\mu v}^\sigma C_{vq}^\sigma + \sum_{j}^{M} 
        \sum_{\mu v \kappa  \theta}^K C_{\mu q}^\sigma C_{ vq}^\sigma (\mu v|\kappa\theta)C_{ \kappa j}^\sigma C_{\theta j}^\sigma  n_j^\sigma\\
        -& \sum_{j}^{M} \sum_{ \mu v \kappa  \theta}^K  C_{\mu q}^\sigma C_{ vj}^\sigma (\mu v|\kappa\theta)C_{ \kappa q}^\sigma C_{\theta j}^\sigma \frac{\partial f(n_q^\sigma ,n_j^\sigma)}{\partial n_q^\sigma}.
    \end{split}\\
    \frac{\partial n_q^\sigma}{\partial x_p^\sigma}=& s_x(x_q^\sigma, \mu^\sigma)   \delta_{pq}+s_\mu(x_q^\sigma, \mu^\sigma) \frac{\partial \mu^\sigma}{\partial x_p^\sigma},
	\end{align}
\end{subequations}
\begin{equation}
    \frac{\partial \mu^\sigma}{\partial x_p^\sigma}=-\frac{s_x(x_p^\sigma, \mu^\sigma)}{V^\sigma},
\end{equation}
and $s_x(x_p^\sigma, \mu^\sigma)=\partial s(x_p^\sigma, \mu^\sigma)/\partial x_p^\sigma$, 
$s_\mu(x_p^\sigma, \mu^\sigma)=\partial s(x_p^\sigma, \mu^\sigma)/\partial \mu^\sigma$, $V^\sigma = \sum_{i=1}^M s_\mu(x_i^\sigma, \mu^\sigma)$. However, the difference between these two variables, $R^\sigma$ and $\{x_p^\sigma\}$, makes the simple first-order algorithms such as gradient descent (GD) and conjugate gradient (CG)\cite{nocedal2006numerical} ineffective. Updating them simultaneously with the same step size and without any modification leads to very slow convergence, even slower than the decoupled optimization methods. In this context, Freysoldt et al. approached the issue by assigning an initial value and adjusting it to find an appropriate preconditioner during the optimization process \cite{freysoldt2009direct}. In contrast, our method utilizes the diagonal elements of the Hessian matrix and their approximations to develop an effective preconditioner.

The second-order algorithm, namely the Newton's method (NM)\cite{nocedal2006numerical}, does not have these difficulties, because it can use the analytical second derivatives to adjust the step size of each variable and achieve fast convergence. The full second derivatives are given below. 

The second derivatives of $E$ with respect to $R_{pq}^\sigma$ are
\begin{equation}
    \begin{split}
      \left.\frac{\partial^2 E}{\partial R_{pq}^\sigma \partial R_{ab}^\sigma}\right|_{R^\sigma=0} = &\delta_{pb}(h_{qa}^\sigma+2J_{qa}^\sigma)(n^\sigma_q-n^\sigma_p+n^\sigma_a-n^\sigma_b)\\
      +&\delta_{pa}(h_{qb}^\sigma+2J_{qb}^\sigma)(n^\sigma_p-n^\sigma_q+n^\sigma_a-n^\sigma_b)\\
      +&\delta_{qb}(h_{pa}^\sigma+2J_{pa}^\sigma)(n^\sigma_q-n^\sigma_p+n^\sigma_b-n^\sigma_a)\\
      +&\delta_{qa}(h_{pb}^\sigma+2J_{pb}^\sigma)(n^\sigma_p-n^\sigma_q+n^\sigma_b-n^\sigma_a)\\
      -&2\delta_{pb} \sum_j^M (qj|aj)[f(n^\sigma_q, n^\sigma_j)-f(n^\sigma_p, n^\sigma_j)\\
      &+f(n^\sigma_a, n^\sigma_j)-f(n^\sigma_b, n^\sigma_j)]\\
      -&2\delta_{pa} \sum_j^M (qj|bj)[f(n^\sigma_p, n^\sigma_j)-f(n^\sigma_q, n^\sigma_j)\\
      &+f(n^\sigma_a, n^\sigma_j)-f(n^\sigma_b, n^\sigma_j)]\\
      -&2\delta_{qb} \sum_j^M  (pj|aj)[f(n^\sigma_q, n^\sigma_j)-f(n^\sigma_p, n^\sigma_j)\\
      &+f(n^\sigma_b, n^\sigma_j)-f(n^\sigma_a, n^\sigma_j)]\\
      -&2\delta_{qa} \sum_j^M (pj|bj)[f(n^\sigma_p, n^\sigma_j)-f(n^\sigma_q, n^\sigma_j)\\
      &+f(n^\sigma_b, n^\sigma_j)-f(n^\sigma_a, n^\sigma_j)]\\
      +&4(n^\sigma_pn^\sigma_a-n^\sigma_pn^\sigma_b-n^\sigma_qn^\sigma_a+n^\sigma_qn^\sigma_b)(pq|ab)\\
       -&4(f(n^\sigma_p,n^\sigma_a)-f(n^\sigma_p,n^\sigma_b)\\
       &-f(n^\sigma_q,n^\sigma_a)+f(n^\sigma_q, n^\sigma_b))(ap|bq),\\
    \end{split}
\end{equation}
where
\begin{equation}
h_{qa}^\sigma = \sum_{\mu v}^K C_{ \mu q}^\sigma h_{\mu v}^\sigma C_{va}^\sigma,
\end{equation}
\begin{equation}
    J_{qa}^\sigma = \sum_{j}^{M} \sum_{\mu v \kappa \theta }^K C_{\mu q}^\sigma C_{va}^\sigma (\mu v|\kappa \theta)C_{\kappa j}^\sigma C_{\theta j}^\sigma n_j^\sigma,
\end{equation}
and
\begin{equation}
    (pq|ab) = \sum_{\mu v \kappa \theta}^K C_{\mu p}^\sigma C_{vq}^\sigma (\mu v|\kappa \theta)C_{\kappa a}^\sigma C_{\theta b}^\sigma.
\end{equation}

The second derivatives of $E$ with respect to $x_p^\sigma$ \cite{Yao2021jpcl,Yao2022jpca} are
\begin{equation}
    \label{eq:EBI2d}
    \begin{split}
    \frac{\partial^2 E}{\partial x_p^\sigma \partial x_q^\sigma} &=\sum_{k=1}^{M} \frac{\partial E}{\partial n_k^\sigma}\frac{\partial^2 n_k^\sigma}{\partial x_p^\sigma \partial x_q^\sigma}\\
    &+\sum_{k,l=1}^{M} \frac{\partial^2 E}{\partial n_k^\sigma \partial n_l^\sigma}\frac{\partial n_k^\sigma}{\partial x_p^\sigma}\frac{\partial n_l^\sigma}{\partial x_q^\sigma},\\
    \end{split}
    \end{equation}
    where
\begin{equation}
\begin{split}
\frac{\partial^2 n_k^\sigma}{\partial x_p^\sigma \partial x_q^\sigma}&= s_{\mu\mu}(x_k^\sigma,\mu^\sigma)\frac{\partial \mu^\sigma}{\partial x_p^\sigma}\frac{\partial \mu^\sigma}{\partial x_q^\sigma}\\
&+s_{x\mu}(x_k^\sigma,\mu^\sigma)(\delta_{kq}\frac{\partial \mu^\sigma}{\partial x_p^\sigma}+\delta_{kp}\frac{\partial \mu^\sigma}{\partial x_q^\sigma})\\
&\!+s_{xx}(x_k^\sigma,\mu^\sigma)\delta_{kq}\delta_{kp} \!+\!s_\mu(x_p^\sigma,\mu^\sigma)\frac{\partial^2 \mu^\sigma}{\partial x_p^\sigma \partial x_q^\sigma},\\
\end{split}
\end{equation}
\begin{equation}
    \begin{split}
        \frac{\partial^2 \mu^\sigma}{\partial x_p^\sigma\partial x_q^\sigma}& =-\frac{1}{V^\sigma}\left[s_{xx}(x_p^\sigma, \mu^\sigma)\delta_{pq}+W^\sigma \frac{\partial \mu^\sigma}{\partial x_q^\sigma}\frac{\partial \mu^\sigma}{\partial x_p^\sigma}\right. \\
        &\left.  s_{x\mu}(x_q^\sigma, \mu^\sigma)\frac{\partial \mu^\sigma}{\partial x_p^\sigma}+s_{x\mu}(x_p^\sigma, \mu^\sigma)\frac{\partial \mu^\sigma}{\partial x_q^\sigma}\right] ,\\
    \end{split}
\end{equation}
and 
$s_{xx}(x_p^\sigma, \mu^\sigma)=\partial^2 s(x_p^\sigma, \mu^\sigma)/\partial x_p^\sigma\partial x_p^\sigma$,
$s_{x\mu}(x_p^\sigma, \mu^\sigma)=\partial^2 s(x_p^\sigma, \mu^\sigma)/\partial x_p^\sigma\partial \mu^\sigma$,
$s_{\mu\mu}(x_p^\sigma, \mu^\sigma)=\partial^2 s(x_p^\sigma, \mu^\sigma)/\partial \mu^\sigma\partial \mu^\sigma$,
$W^\sigma = \sum_{i=1}^M s_{\mu\mu}(x_i^\sigma, \mu^\sigma)$.

The second derivatives with respect to both $R_{pq}^\sigma$ and $x_a^\sigma$ are also required, and they are
\begin{equation}
\left.\frac{\partial^2 E}{\partial R_{pq}^\sigma \partial x_b^\sigma}\right|_{R^\sigma=0}=\sum_a^M \frac{\partial^2 E}{\partial R_{pq}^\sigma \partial n_a^\sigma}\frac{\partial n_a^\sigma}{\partial x_b^\sigma},
\end{equation}
where
  \begin{equation}
    \begin{split}
        \left.\frac{\partial^2 E}{\partial R_{pq}^\sigma \partial n_a^\sigma}\right|_{R^\sigma=0} &= \delta_{aq} \left[2h_{pq}^\sigma\!+\!2\sum_{j}^{M} (pq|jj) n_j^\sigma-2\sum_{j}^{M} (pj|qj)\frac{\partial f(n_q, n_j)}{\partial n_q}\right]\\
        &+\left[2(pq|aa)n_q-2(pa|qa)\frac{\partial f(n_a, n_q)}{\partial n_a}\right]\\
        &-\delta_{ap} \left[2h_{pq}^\sigma\!+\!2\sum_{j}^{M} (pq|jj) n_j^\sigma-2\sum_{j}^{M} (pj|qj)\frac{\partial f(n_p, n_j)}{\partial n_p}\right]\\
        &-\left[2(pq|aa)n_p-2(pa|qa)\frac{\partial f(n_a, n_p)}{\partial n_a}\right]\\
    \end{split}
  \end{equation}

The second-order algorithm has a fast convergence, but it is impractical for large systems due to the complexity of the formulae. Therefore, we stick to the first-order algorithms for coupled optimization and introduce some additional enhancements to achieve faster and more efficient convergence. Indeed, the concept of simultaneous optimization of orbitals and occupations has been applied in DFT, such as in finite temperature DFT \cite{freysoldt2009direct}. However, in RDMFT, most approaches have traditionally followed a decoupled scheme. Recent work by Cartier et al.  introduced a one-step procedure to simultaneously optimize NOs and ONs\cite{Cartier2024jctc}, showing that coupled optimization yields superior results compared to decoupled methods. While our approach aligns with theirs in using the EBI method for managing the ON component, a fundamental difference is evident in the handling of the Hessian matrix. Our work focuses on utilizing the diagonal elements of the Hessian matrix as preconditioners for the gradient, contrasting with their approach of approximating the Hessian matrix. Furthermore, we have developed an efficient line search strategy to significantly enhance the speed of convergence.

\subsection{Preconditioner and line search} 

To obtain better convergence in coupled optimization using the first-order algorithm, we apply preconditioned conjugate gradient (PCG) with line search \cite{nocedal2006numerical}. Algorithm \ref{alg: simultaneous} illustrates the algorithm flow we devise.

\begin{algorithm}
    \caption{Coupled optimization of unitary and EBI}\label{alg: simultaneous}
    \SetInd{0.5em}{1em}
    \SetNlSkip{0.5em}
    \SetNlSty{textbf}{}{:}
    \SetAlgoNoLine
     Generete  initial guesses for NOs and ONs: $C^\sigma_0$ and $n^\sigma_0$; calculate $x_0^\sigma$ corresponding to $n^\sigma_0$; set $R^\sigma_0=0$, $\Delta E = 1$, $k=1$\;
     Calculate initial  gradient $g_{{R}_0} = \partial E/\partial {R}^\sigma_0$, $g_{x_0}= \partial E/\partial x_0^\sigma$, and preconditioner $P_{{R}_0}$ and $P_{x_0}$ and energy $E_0$\;
      Set $p_{{R}_0}=0, p_{x_0}=0,z_{{R}_0}=1, z_{x_0}=1$\;
     \While{$\mathrm{norm}(|g_{{R}_k}|)>1e-4$ and $\mathrm{norm}(|g_{x_k}|)>1e-4$ and $|\Delta E|>1e-8$ }{
        Calculate $z_{{R}_k}=g_{{R}_k}/P_{{R}_k}$, $z_{x_k}=g_{x_k}/P_{x_k}$\; 
        Calculate $\beta_{{R}_k}=(g_{{R}_k}^T(z_{{R}_k}-z_{{R}_{k-1}}))/(g_{{R}_{k-1}}^Tz_{{R}_{k-1}}) $, $\beta_{x_k}=(g_{x_k}^T(z_{x_k}-z_{x_{k-1}}))/(g_{x_{k-1}}^Tz_{x_{k-1}})$\;
        \If {$\mathrm{abs}(p_{{R}_{k-1}}^T g_{{R}_k})>0.2*p_{{R}_{k-1}}^T g_{{R}_{k-1}}$}{$\beta_{{R}_k}=0$\;}
        \If {$\mathrm{abs}(p_{x_{k-1}}^T g_{x_k})>0.2*p_{x_{k-1}}^T g_{x_{k-1}}$}{$\beta_{x_k}=0$\;}
        $p_{{R}_k}= z_{{R}_{k}}+\beta_{{R}_k}p_{{R}_{k-1}}$, $p_{x_k}= z_{x_{k}}+\beta_{x_k}p_{x_{k-1}}$ \;
        Line search  to obtain the optimal step size, $\alpha_{{R}_k}$ and$\alpha_{x_k}$\; 
        Update $C_{k+1}^\sigma = C_{k}^\sigma e^{-\alpha_{{R}_k}p_{{R}_k}}$, $x_{k+1}^\sigma = x_{k}^\sigma -\alpha_{x_k}p_{x_k}$\;
        Calculate  $E_{k+1}$, $g_{{R}_{k+1}}$, $g_{x_{k+1}}$, and preconditioner $P_{{R}_{k+1}}$ and $P_{x_{k+1}}$ \;
        Calculate $\Delta E =E_{k+1}-E_k$\; 
     $k\leftarrow k+1$\;}
\end{algorithm}

In Algorithm \ref{alg: simultaneous}, the preconditioners, $P_{{R}}$ and $P_{x}$, and the line search are essential for the efficiency of this algorithm. Here, we derive the preconditioner from the diagonal element of the second derivatives.
The diagonal elements of the second derivatives with respect to ${R}_{pq}^\sigma$ are
\begin{equation}
    \begin{split}
      &  \left.\frac{\partial^2 E}{\partial {R}_{pq}^{\sigma }\partial {R}_{pq}^{\sigma }} \right|_{{R}^\sigma=0}=(2h_{pp}^\sigma+4J_{pp}^\sigma-2h_{qq}^\sigma-4J_{qq}^\sigma)(n^\sigma_q-n^\sigma_p)\\
      &-4\sum_j^{M} [(pj|pj)-(qj|qj)][f(n^\sigma_q, n^\sigma_j)-f(n^\sigma_p, n^\sigma_j)]\\
      &+4(n_p^\sigma- n_q^\sigma )^2(pq|pq)\\
      &-4[f(n^\sigma_p,n^\sigma_p)+f(n^\sigma_q,n^\sigma_q)
-f(n^\sigma_p,n^\sigma_q)\\
       &\ -f(n^\sigma_q, n^\sigma_p)]\times
       (pp|qq)\\
    \end{split}
  \end{equation}
However, the computation cost of the above formula is still high. Thus, we simplify it by omitting the last two terms, and obtain the preconditioner for ${R}_{pq}^\sigma$ as
\begin{equation}
    \begin{split}
      &P_{{R}^\sigma} =(2h_{pp}^\sigma+4J_{pp}^\sigma-2h_{qq}^\sigma-4J_{qq}^\sigma)(n^\sigma_q-n^\sigma_p)\\
      &-4\sum_j^{M} [(pj|pj)-(qj|qj)][f(n^\sigma_q, n^\sigma_j)-f(n^\sigma_p, n^\sigma_j)]\\ \label{p orb}
    \end{split}
  \end{equation}
The preconditioner is very easy to compute and has a good effect. 
In order to guarantee the positive definite of the preconditioner, the $P_{{R}^\sigma}$ calculated by Eq. \ref{p orb} is processed as follows in the program: If the minimum value of $P_{{R}^\sigma}$ is negative, set $P_{{R}^\sigma}=P_{{R}^\sigma}-\min (P_{{R}^\sigma})$. 
In order to avoid numerical problems due to $P_{{R}^\sigma}$ being used as the denominator, a threshold is set to 0.00001, and the elements of $P_{{R}^\sigma}$ that are less than the threshold are set equal to the threshold.

The diagonal elements of the second derivatives with respect to $x_p^\sigma$ are
\begin{equation}
\begin{split}
\frac{\partial^2 E}{\partial x_p^{\sigma } \partial x_p^{\sigma }}&=\sum_{k=1}^{M} \frac{\partial E}{\partial n_k^\sigma}\frac{\partial^2 n_k^\sigma}{\partial x_p^\sigma \partial x_p^\sigma}\\
&+\sum_{k,l=1}^{M}\frac{\partial^2 E}{\partial n_k^\sigma \partial n_l^\sigma}\frac{\partial n_k^\sigma}{\partial x_p^\sigma}\frac{\partial n_l^\sigma}{\partial x_p^\sigma},\\
\end{split}
\end{equation}
By omitting the second term, we obtain a simple preconditioner for $x_p^\sigma$ 
\begin{equation}
\label{eq:p1x}
P_{x_p^\sigma}^1 = \sum_{k=1}^{M} \frac{\partial E}{\partial n_k^\sigma}\frac{\partial^2 n_k^\sigma}{\partial x_p^\sigma \partial x_p^\sigma}.
\end{equation}
To achieve better convergence, the preconditioner for $x_p^\sigma$ that combines Eq. \ref{eq:p1x} with the Broyden-Fletcher-Goldfarb-Shanno (BFGS) approximation \cite{broyden1970convergence, fletcher1970new, goldfarb1970family, shanno1970conditioning} is utilized here, which is
\begin{equation}
P_{x_i^\sigma} = g P_{x_p^\sigma}^{BFGS}+(1-g)P_{x_p^\sigma}^1,
\end{equation}
where the BFGS preconditioner is
\begin{equation}
P_{x_p^\sigma}^{BFGS} = (B_{k+1})_{pp},
\end{equation}
with
\begin{equation}
B_{k+1}=B_{k}+\frac{y_ky_k^T}{y_k^Ts_k}-\frac{B_ks_ks_k^TB_k^T}{s_k^TB_ks_k},
\end{equation}
and $s_k=x_{k+1}-x_k$, $y_k=\partial E/\partial x_{k+1}-\partial E/\partial x_{k}$.  $g$ is a hybridization parameter with a value range between 0 and 1. According to the test results, the recommended value of $g$ is 0.9.

Line search can be performed by using a quadratic function approximation \cite{Traian2009} to find the optimal step size. By expanding the energy with respect to the step size $\alpha$ up to second order, we have
\begin{equation}
E(\alpha) \approx  a \alpha^2+b\alpha+c
\end{equation}
The optimal step size can be obtained from the minimum point, which is
\begin{equation}
\label{eq:alpha}
\alpha = \frac{E'(0)\tilde{\alpha}}{E'(0)-E'(\tilde{\alpha})}
\end{equation}
where $\tilde{\alpha}$  is a trial step size between 0 and 1. Therefore, to obtain the optimal step size, we also need to compute the derivative $E'(\tilde{\alpha})$ at the trial step size $\tilde{\alpha}$. Alternatively, we can also expand the energy with respect to both step sizes as
\begin{equation}
\label{eq:a12}
E(\alpha_{R},\alpha_x ) \approx  a_{1} \alpha_{R}^2+a_{2}\alpha_x^2+ a_{3} \alpha_{{R}}\alpha_{x}+b_{1}\alpha_H+b_{2}\alpha_x+c
\end{equation}
and minimize it to obtain the optimal step sizes $\alpha_{{R}}$ and $\alpha_{x}$. In this work, we use the simple Eq. \ref{eq:alpha} to obtain the step sizes $\alpha_{{R}}$ and $\alpha_{x}$ for ${R}^\sigma$ and $x^\sigma$, respectively. To simplify the computation, we determine $\alpha_{{R}}$ and $\alpha_{x}$ together, which means we only need one extra computation, with trial step sizes $\tilde{\alpha}_{R}$ and $\tilde{\alpha}_x$, to obtain the derivatives $E'(\tilde{\alpha})$ required for calculating the optimal step sizes $\alpha_{{R}}$ and $\alpha_{x}$. In other words, we ignore the term $a_{3} \alpha_{{R}}\alpha_{x}$ in Eq. \ref{eq:a12} when determining the step sizes, and this can already achieve a very good effect.

It should be noted that the coupled optimization method proposed above has a great computational benefit, because it optimizes NOs and ONs simultaneously in one step with almost the same computational cost as optimizing NOs or ONs individually in one step. This benefit, together with better convergence speed, makes the coupled method have a clear edge over the decoupled method.

\section{Computational Details \label{comp}}

The coupled optimization (CO) method is evaluated on various functionals, basis sets, and systems. 
 To compare its performance, we also check some of the decoupled optimization methods, where NOs are unitarily optimized using PCG with the same preconditioner as that in the CO method, while ONs are optimized using different methods and algorithms. These methods include ALM and EBI, and these algorithms include CG, PCG, and NM. The tested decoupled methods are denoted as ALM@CG, ALM@NM, EBI@PCG, EBI@CG, and EBI@NM, respectively. Note that the CO method optimizes NO and ON simultaneously, so the NO and ON iterations are both equal to the overall iterations, while the decoupled methods have total iterations that are the sum of the NO and ON iterations. In the following tests, the CO method of Algorithm \ref{alg: simultaneous} is denoted as EBI@CO. All calculations were performed using a local software package, which called the LIBINT integral library \cite{valeev2020libint} and the  LIBXC library \cite{lehtola2018recent}. The convergence criteria for energy and gradient were set to $10^{-8}$ and $10^{-4}$ a.u., respectively. A calculation is considered converged when both the energy and gradient criteria are satisfied simultaneously.

 For the sake of testing the robustness of different optimization methods, all methods utilize the same initial setup as described here. For the orbital coefficients $C^\sigma$, the superposition of atomic densities (SAD) method \cite{almlof1982principles,van2006starting} is used to obtain the initial density matrix. This is a common method for generating initial guesses of the density matrix in many computational programs. Following this, the Fock matrix is calculated based on the initial density matrix. Diagonalization of the Fock matrix provides the initial orbital coefficients and orbital energies. Note that various initial guesses, including the SAD, have been recently assessed \cite{Lehtola2019jctc}. The procedure for generating initial guesses for the occupations $\{n^\sigma_p\}$ is summarized in Algorithm \ref{alg: occ}, where the initial occupations exhibit the following characteristics: the occupations of occupied orbitals are close to 1, the occupations of unoccupied orbitals are close to 0, and orbitals with lower energy have larger occupations. Therefore, such initial guesses for orbitals and occupations are generally reasonable and effective.
 
 \begin{algorithm}
   \caption{Initial guess generation for occupations}\label{alg: occ}
   \SetInd{0.5em}{1em}
   \SetNlSkip{0.5em}
   \SetNlSty{textbf}{}{:}
   \SetAlgoNoLine
   Initialize $x_p^\sigma=2$ for $p=1,...,N^\sigma$, and $x_p^\sigma=-2$ for $p=N^\sigma+1,...,{M}$\;
    Solve the implicit function in EBI and obtain the corresponding occupations $\{n_p^\sigma\}$ that satisfies the constraints\;
   Sort the occupations $\{n_p^\sigma\}$ according to orbital energies so that orbitals with lower energies have larger occupations.\;
    \Return{$\{n_p^\sigma\}$\;}
 \end{algorithm}

 In KS-DFT, molecular orbitals typically assume integer occupations. However, in RDMFT, the dispersion of occupations varies significantly across different functionals, making it challenging to provide a universally optimal initial guess. Furthermore, the optimal orbitals corresponding to these varying occupations also differ, complicating the reliability of any initial guess for orbitals.
 
 To address these challenges, our approach involves testing the dependence of optimization algorithms on different initial conditions. To generate a range of initial guesses and assess this dependence, we introduce random perturbations to reasonable initial guesses. 
 
 The initial orbital coefficients are first derived using SAD. These coefficients are further perturbed to test the robustness of our optimization methods under diverse starting conditions (see Algorithm \ref{alg: orb random}). Initial guesses for occupations are also modified with random perturbations to evaluate the stability and performance of different optimization methods under different initial conditions (see Algorithm \ref{alg: occ random}).

 \begin{algorithm}
   \caption{Initial guess generation for orbitals with random perturbation}\label{alg: orb random}
   \SetInd{0.5em}{1em}
   \SetNlSkip{0.5em}
   \SetNlSty{textbf}{}{:}
   \SetAlgoNoLine
 
   Generate  orbital coefficients $C^\sigma$ via SAD  \;
   Generate an $M \times M$ random matrix ${R}^\sigma$ with all elements uniformly distributed within the range [0,1] ;
   Adjust ${R}^\sigma$ to be anti-symmetric:  ${R}^\sigma \rightarrow [({R}^\sigma)^T-{R}^\sigma]*0.1$ \;
   Apply the exponential map to introduce perturbations: $C^\sigma \rightarrow C^\sigma e^{{R}^\sigma}$\;
    \Return{$C^\sigma$}
 \end{algorithm}
 
 \begin{algorithm}
   \caption{Initial guess generation for occupations with random perturbation}\label{alg: occ random}
   \SetInd{0.5em}{1em}
   \SetNlSkip{0.5em}
   \SetNlSty{textbf}{}{:}
   \SetAlgoNoLine
   Assign random values uniformly distributed within [0.5, 1] for occupied orbitals $n_p^\sigma$ for $p=1,...,N^\sigma$ \;
 Normalize the remaining occupations: $n_p^\sigma \leftarrow (N^\sigma - \sum_{q=1}^{N^\sigma} n_q^\sigma) / (M - N^\sigma)$ for $p=N^\sigma+1,...,M$ \;
 Order occupation numbers according to orbital energies, ensuring lower energy orbitals have higher occupations \;
 \Return{${n_p^\sigma}$}
 \end{algorithm}

To examine the convergence and applicability of EBI@CO for different 1-RDM functionals, we performed the following tests. First, taking $\mathrm{C_6H_6}$ as example, three different basis sets: 6-31G \cite{hehre1972self}, cc-pVDZ, and cc-pVTZ \cite{kendall1992electron, woon1993gaussian} were utilized to investigate the influence of basis set size on the convergence. Second, the power functionals\cite{Sharma2008prb} of the form $f(n_p^\sigma, n_q^\sigma) = (n_p^\sigma n_q^\sigma)^m$ , with $m$ varying from 0.1 to 0.9 were employed to assess the robustness of the optimization methods for different functionals. Third, initial guesses with random perturbation of NOs and ONs were used to access the stability of the optimization methods. 

Next, 70  systems were calculated, covering a wide range of chemical scenarios. These systems included: 10 transition metal complexes ($\mathrm{Co(H_2O)_6}$, $\mathrm{Cr(CO)_5(C_2H_4)}$, $\mathrm{Cr(CO)_5(H_2)}$, $\mathrm{Cr(CO)_6}$, $\mathrm{Cu(NH_3)_6}$, $\mathrm{Fe(CO)_5}$,  $\mathrm{Fe(NH_3)_6}$,  $\mathrm{Mn(CO)_6}$,  $\mathrm{Ni(H_2O)_6}$, $\mathrm{Ti(CO)_6}$), 10 free radicals ($\mathrm{C_1H_1}$, $\mathrm{C_1H_3}$, $\mathrm{C_1N_1}$,  $\mathrm{C_1O_1H_1}$, $\mathrm{C_1O_1H_3}$,  $\mathrm{C_1S_1H_3}$, $\mathrm{C_2H_1}$, $\mathrm{C_2H_3}$, $\mathrm{C_3H_7}$, $\mathrm{O_1H_1}$), 10 weak interaction systems ($\mathrm{C_2H_4}$ dimer, $\mathrm{C_6H_6\!-\!Ne}$, $\mathrm{CH_3SH\!-\!HCl}$, $\mathrm{CH_4\!-\!Ne}$, $\mathrm{H_2S\!-\!H_2S}$, $\mathrm{HCONH_2}$ dimer, $\mathrm{He\!-\!Ar}$, $\mathrm{HF}$ dimer, $\mathrm{NH_3\!-\!ClF}$, $\mathrm{C_6H_6}$ dimer), 10 molecules containing halogens ($\mathrm{CClH_3}$, $\mathrm{CCl_2H_2}$, $\mathrm{CCl_4}$, $\mathrm{CF_2H_2}$, $\mathrm{CF_4}$, $\mathrm{CHCl_3}$, $\mathrm{CHF_3}$, $\mathrm{C_2ClH_3}$, $\mathrm{C_2ClH_5}$, $\mathrm{C_2F_4}$), 10 molecules containing metal elements ($\mathrm{AlCl_3}$, $\mathrm{AlF_3}$, $\mathrm{KOH}$, $\mathrm{LiF}$, $\mathrm{LiH}$, $\mathrm{Li_2}$, $\mathrm{LiOH}$, $\mathrm{MgO}$, $\mathrm{NaCl}$, $\mathrm{Na_2}$), and 20 molecules containing C, O, N, P,  S and Si elements ($\mathrm{CSiH_6}$, $\mathrm{SO}$, $\mathrm{PH_3}$, $\mathrm{SH_2}$, $\mathrm{SO_2}$, $\mathrm{S_2}$, $\mathrm{CS_2}$, $\mathrm{SiH_4}$, $\mathrm{Si_2}$, $\mathrm{Si_2H_6}$, $\mathrm{CNH}$, $\mathrm{CNH_5}$, $\mathrm{CNO_2H_3}$, $\mathrm{COH_4}$, $\mathrm{C_2H_2O}$, $\mathrm{C_2NH_7}$, $\mathrm{C_2OH_4}$, $\mathrm{C_2OH_6}$, $\mathrm{N_2C_2}$, $\mathrm{COH_2}$). Different basis sets are employed based on the system's characteristics: the aug-cc-pVDZ set for 10 molecules involving weak interactions, def2-TZVP  \cite{weigend2005balanced} for 10 systems containing metal elements, def2-SVP \cite{weigend2005balanced} for 10 transition metal complexes, and cc-pVTZ for the remaining 40 molecules.

Finally, a series of increasing olefin chains were calculated to investigate the scalability of the mehods as the system size increases, and the $\mathrm{C_{60}}$ system was tested to demonstrate the ability of EBI@CO for large systems, with the basis set cc-pVDZ.

\section{Results and Discussions \label{results}}

In this study, we conduct a comprehensive comparison of different optimization methods using the $\omega$P22  \cite{LYPr,Ai2022jpcl,Ai2023jcp} and the power functionals \cite{Sharma2008prb}, across various basis sets. The effectiveness of each method is evaluated in terms of convergence speed, converged energies, and sensitivity to basis set size, with findings illustrated in the corresponding figures.

\subsection{Evaluation of Optimizations Across Various Functionals and Basis Sets}

\begin{figure}[t]
    \includegraphics[width=0.7\linewidth]{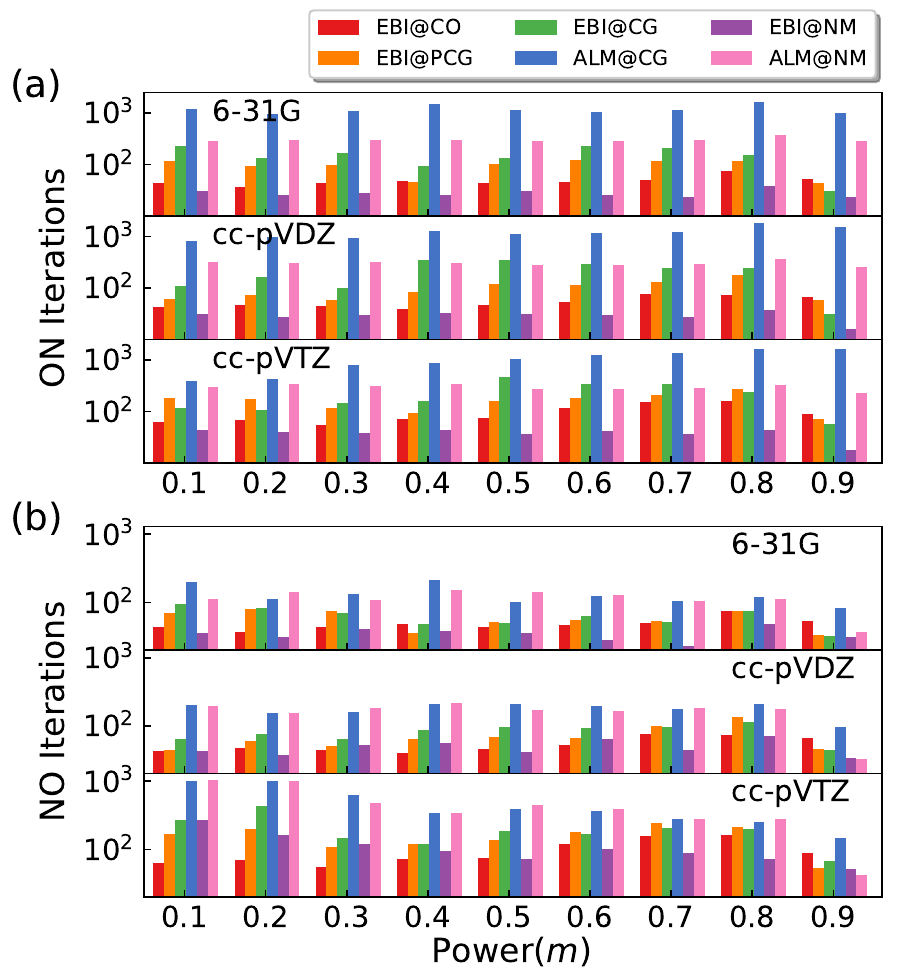}
    \caption{Comparison of the number of iterations required for different optimization methods to optimize (a) ONs and (b) NOs of $\mathrm{C_6H_6}$ using power functionals with $m \in [0.1, 0.9]$ and various basis sets. Logarithmic $y$-axis is used.}
    \label{fig:basis_nocc_norb}
    \end{figure}

The overall iterations of ONs and NOs are first compared for different optimization methods applied to $\mathrm{C_6H_6}$ using power functionals with $m$ ranging from 0.1 to 0.9 and different basis sets. The results are shown in Fig. \ref{fig:basis_nocc_norb}. 
The EBI@CO method demonstrates significantly faster convergence than the decoupled methods, independent of both the choice of basis set and the functional parameter $m$. Furthermore, EBI@CO exhibits minimal sensitivity to the size of the basis set, showing a substantially smaller increase in iteration counts compared to its decoupled counterparts. This robust performance can be attributed to its integrated optimization process, which circumvents the synchronization issues encountered in separate optimizations of NOs and ONs. In contrast, decoupled methods suffer from asynchronous updates between NOs and ONs, leading to mutual interference and delayed convergence.

    \begin{figure}
    \includegraphics[width=0.65\linewidth]{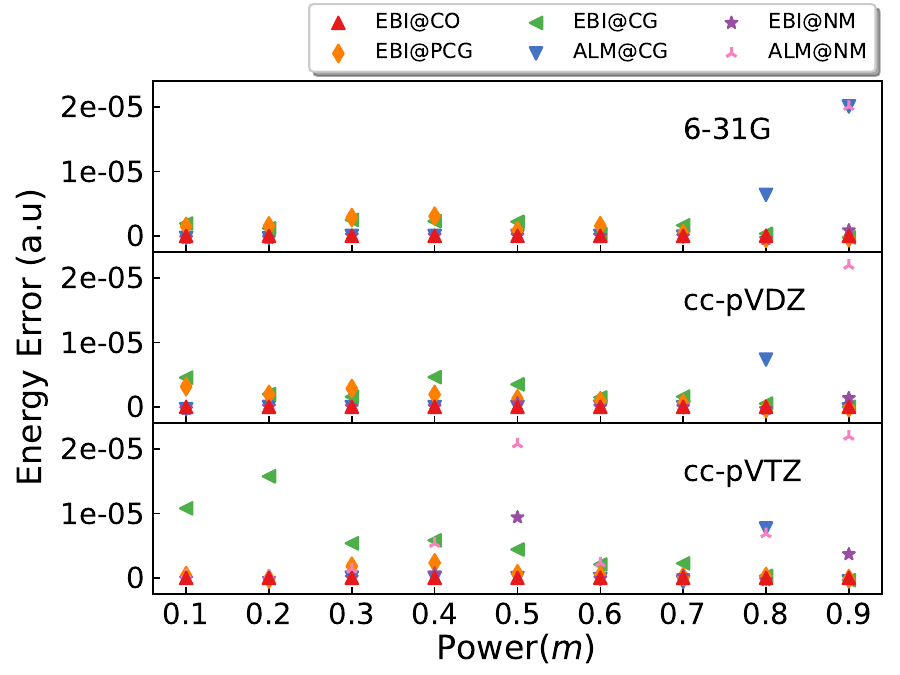}
    \caption{Comparison of energy errors for different optimization methods and basis sets applied to $\mathrm{C_6H_6}$ using power functionals with $m \in [0.1, 0.9]$. The energies obtained using the EBI@CO method serve as the reference. Energy errors for other methods are calculated relative to EBI@CO, expressed as $\Delta E=E$(method)$-E$(EBI@CO).}
    \label{fig:basis_energy}
    \end{figure}

Fig. \ref{fig:basis_energy} further shows energy errors during optimization for different methods using power functionals with $m$ ranging from 0.1 to 0.9 and different basis sets. The energies at convergence of EBI@CO are set to 0. These results indicate that EBI@CO not only converges much faster than the decoupled methods, but also has lower converged energies than them. The reason why decoupled optimization frequently encounters problems with local minima can be attributed to the asynchronous optimization of NOs and ONs. This desynchronization may prevent effective navigation of the energy landscape, increasing the risk of getting stuck in suboptimal minima.

The decoupled methods are examined in detail to understand their performance limitations. LM is notably affected by the initial guess's quality, with its convergence reliability decreasing as the basis set expands. For some functionals with the larger basis set cc-pVTZ, achieving convergence proved elusive even after exploring 20 different initial guesses. ALM, while more robust than LM, demands numerous iterations to satisfy the constraints for ONs, substantially increasing the computational effort required. Among the decoupled methods, the EBI@PCG emerges as the superior first-order method, thanks to the implementation of a novel preconditioner, which enhances the optimization process, enabling EBI@PCG to outperform other first-order methods.

\begin{figure}[t]
\includegraphics[width=0.65\linewidth]{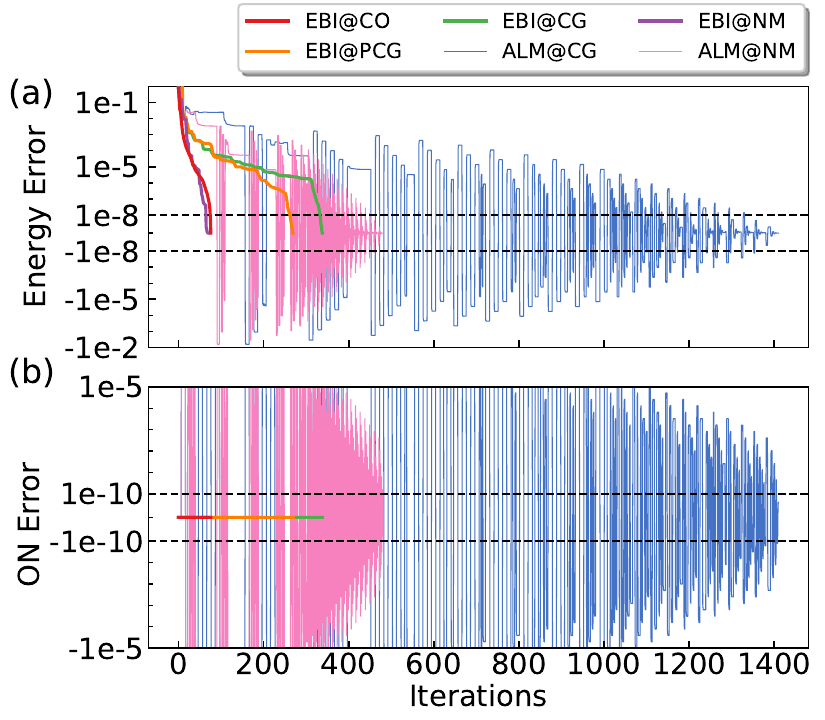}
\caption{Comparison of energy and ON errors for different optimization methods applied to $\mathrm{C_6H_6}$ using the power functional with $m=0.7$ and the cc-pVDZ basis set. (a) Energy errors versus optimization iterations. The energies obtained using the EBI@CO method serve as the reference. Energy errors for other methods are calculated relative to EBI@CO, expressed as $\Delta E=E$(method)$-E$(EBI@CO). The energy units are a.u. (b) ON errors versus optimization iterations. ON errors are quantified by the expression $\sum_\sigma(\sum_i n_i^\sigma - N_0^\sigma)$.}
\label{fig:energy_process_p07}
\end{figure}

Building on the comparative analysis of optimization methods, we delve into the specifics of energy and ON errors to elucidate the convergence behaviors exhibited by these methods. Fig. \ref{fig:energy_process_p07} illustrates the optimization procedure for benzene ($\mathrm{C_6H_6}$) using the power functional with $m=0.7$ and the cc-pVDZ basis set. Figure 3(a) reveals a direct correlation between the energy accuracy in the ALM method and the convergence associated with ONs. The optimization strategy employed by ALM, as seen in Eq. \ref{eq:ALM}, leads to continual fluctuations in the total occupations, which in turn causes oscillations in energy. Notably, the EBI@CO method, leveraging a first-order algorithm, achieves convergence with an efficacy comparable to that of EBI@NM, underscoring the effectiveness of coupled optimization strategies. This observation demonstrates the benefit of the coupled method, offering insights into the potential advantages of integrating ON and NO optimizations to enhance the convergence and accuracy of RDMFT calculations.

\begin{table}[ht]
\caption{Average iterations and average energy errors of EBI@CO and different decoupled optimization methods for $\mathrm{C_6H_6}$ with different functionals and basis sets. Energy errors are calculated with respect to the reference values obtained by EBI@CO, where a negative energy error indicates that a method has achieved a lower energy than EBI@CO. The unit of energy is a.u. VDZ and VTZ refer to the cc-pVDZ and cc-pVTZ basis sets, respectively.}
    \begin{tabular}{ccccc}
        \hline
        \hline
        \multirow{2}{*}{Algorithms} \quad &  \multirow{2}{*}{Basis set} \quad & \quad \multirow{2}{*}{\makecell[c]{Energy \\ error}} \quad  & \qquad \multirow{2}{*} {\makecell[c]{NO \\ iter.}} \qquad\quad &  \multirow{2}{*}{\makecell[c]{ON \\ iter.}}   \qquad \\
    \\
    \hline 
    \hline 
    \multirow{3}{*}{EBI@CO}     &  6-31G    & -          &   49.00     &  49.00       \\
    & VDZ       & -          &   54.56     &  54.56       \\
    & VTZ       & -          &   95.78     &  95.78       \\			
\hline 			
\multirow{3}{*}{EBI@PCG}    &  6-31G    &  1.21$\times10^{-6}$  &   59.11 	&  95.67       \\
    & VDZ       &  1.35$\times10^{-6}$  &   70.89 	&  98.78       \\
    & VTZ       &  6.21$\times10^{-7}$  &   157.67 	&  163.44       \\
\hline 	
\multirow{3}{*}{EBI@CG}     &  6-31G    &  1.34$\times10^{-6}$  &   63.22 	&  155.22       \\
    & VDZ       &  2.20$\times10^{-6}$  &   81.89 	&  209.78       \\
    & VTZ       &  5.15$\times10^{-6}$  &   200.56 	&  220.67       \\
\hline 
\multirow{3}{*}{ALM@CG}     &  6-31G    &  2.84$\times10^{-6}$  &   133.00 	&  1184.22      \\
    & VDZ       &  7.11$\times10^{-7}$  &   180.44 	&  1213.44      \\
    & VTZ       &  8.94$\times10^{-7}$  &   492.33 	&  1067.44      \\
\hline 
\multirow{3}{*}{EBI@NM}     &  6-31G    &  -4.76$\times10^{-8}$ &   34.56 	&  28.22       \\
    & VDZ       &  1.18$\times10^{-7}$  &   49.22 	&  29.22       \\
    & VTZ       &  1.46$\times10^{-6}$  &   115.78 	&  37.89       \\
\hline 
\multirow{3}{*}{ALM@NM}     &  6-31G    &  2.08$\times10^{-6}$  &   115.67 	&  305.00       \\
    & VDZ       &  2.36$\times10^{-6}$  &   164.00 	&  305.11       \\
    & VTZ       &  6.53$\times10^{-6}$  &   478.44 	&  300.67       \\
    \hline 
    \hline 
    \end{tabular} \label{tab:basis}
\end{table}

Table \ref{tab:basis} summarizes the performance of different optimization methods, focusing on the average iterations required and the energy errors observed. Analysis of the table reveals that EBI@PCG demonstrates a remarkable consistency in the average number of iterations for optimizing NOs and ONs. This consistency is indicative of uniform accuracy throughout the optimization process, which significantly contributes to EBI@PCG's expedited convergence. Such efficient convergence of EBI@PCG, attributed to its consistent optimization of NOs and ONs, showcases the effectiveness of the preconditioner employed. Notably, EBI@CO, which utilizes the same preconditioner, also exhibits commendable convergence performance. This underscores the preconditioner's role in enhancing the optimization process for both coupled and decoupled methods. These findings emphasize the critical importance of selecting an optimization method that ensures consistency and accuracy across both NOs and ONs. The demonstrated efficiency of the preconditioner in both EBI@PCG and EBI@CO suggests a promising direction for future optimization strategies.

\subsection{Impact of Initial Guess Variability on Optimization Performance}

\begin{figure}[t]
\includegraphics[width=0.6\linewidth]{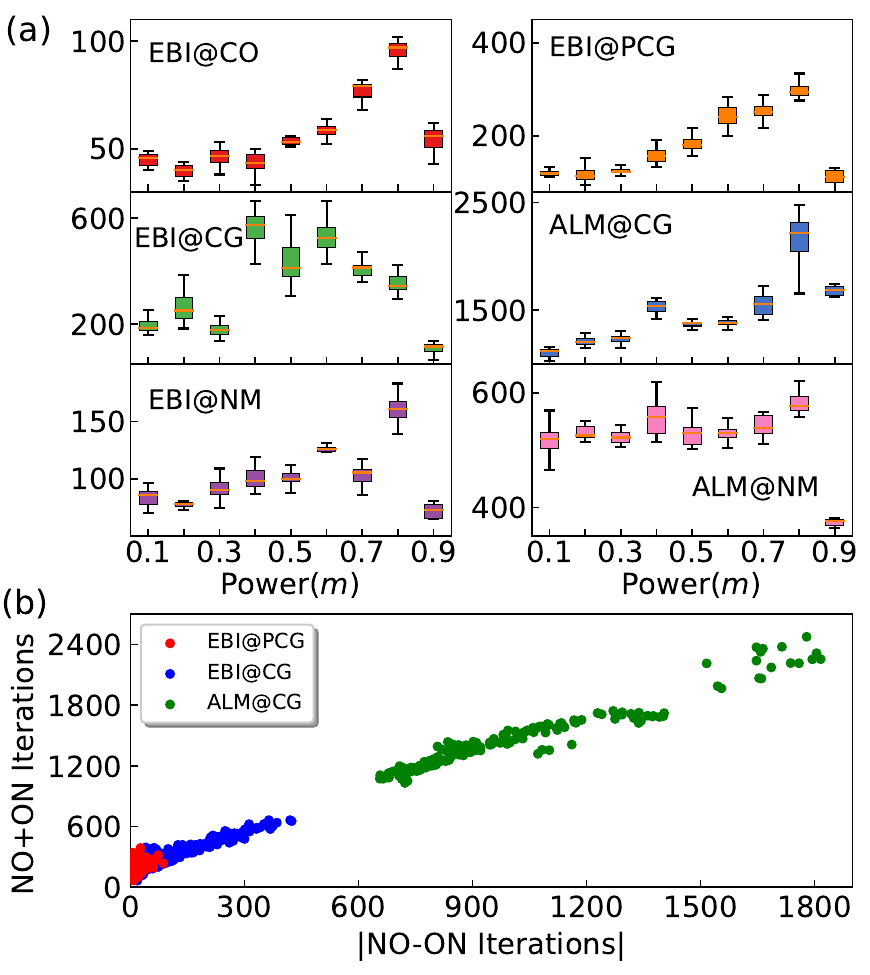}
\caption{Number of iterations of EBI@CO and different decoupled methods for $\mathrm{C_6H_6}$ using 20 initial guesses with random perturbation. The power functionals with $m \in [0.1, 0.9]$ and the cc-pVDZ basis set are used. (a) Box plot of the iterations for 20 initial guesses. (b) Scatter plot of the total iterations versus the difference of the NO and ON iterations for EBI@PCG, EBI@CG and ALM@CG. }
\label{fig:niter}
\end{figure}

\begin{table}[!ht]
\caption{Average iterations and energy errors of EBI@CO and different decoupled methods for $\mathrm{C_6H_6}$ using 20 initial guesses with random perturbation. The power functionals with $m \in [0.1, 0.9]$ and the cc-pVDZ basis set are used. The lowest energy obtained using the EBI@CO method serves as the reference for energy error calculation of each method. The unit for energy is a.u.}
        \begin{tabular}{ccccc}
            \hline
            \hline
             \multirow{2}{*}{Algorithms} \quad \quad&    \multirow{2}{*} {\makecell[c]{NO \\ iter.}} \quad \quad&   \multirow{2}{*}{\makecell[c]{ON \\ iter.}} \qquad  & \multirow{2}{*}{\makecell[c]{Sum \\ iter.}} \qquad & \multirow{2}{*}{\makecell[c]{Energy \\ error}} \quad   \\
        \\
        \hline 
        \hline 
        EBI@CO            &   56.88 	   &      56.88 	    &    56.88   	&  1.82$\times10^{-7}$   \\
        EBI@PCG            &   83.87 	   &      95.46 	    &    179.33 	&  1.20$\times10^{-6}$   \\
        EBI@CG            &   101.26 	 &      236.05 	    &    337.31 	&  2.29$\times10^{-6}$   \\
        ALM@CG            &   228.35 	 &      1233.51 	&    1461.86 	&  1.07$\times10^{-6}$   \\
        EBI@NM            &   68.79 	   &      33.03 	    &    101.83 	&  2.32$\times10^{-7}$   \\
        ALM@NM            &   214.71 	 &      305.66 	    &    520.37 	&  2.44$\times10^{-6}$   \\
        \hline 
        \hline 
        \end{tabular} \label{tab:rand}
\end{table}

To assess the stability and reliability of various optimization methods, we performed an evaluation using 20 initial guesses with random perturbation for each power functional within the range $m \in [0.1, 0.9]$, focusing on convergence speed and energy accuracy for the $\mathrm{C_6H_6}$ molecule with the cc-pVDZ basis set. As depicted in Fig. \ref{fig:niter}a, the EBI@CO method demonstrates remarkable stability, consistently converging to an accuracy of $10^{-8}$ within just 100 iterations across all tested initial guesses. Fig. \ref{fig:niter}b illustrates a clear correlation between the total number of iterations required for convergence and the discrepancy in iterations between optimizing NOs and ONs. This trend highlights the necessity for achieving synchronous accuracy in NO and ON optimizations to ensure efficient and simultaneous convergence. Table \ref{tab:rand} summarizes the performance for all methods based on random initial guesses. Notably, the EBI@CO method outperforms decoupled methods, averaging 56.88 iterations with an exceptional energy accuracy of $10^{-7}$. This starkly contrasts with the higher iteration counts and less consistent energy accuracies observed for decoupled methods, underlining EBI@CO's superior stability and robustness. These evaluations underscore the critical importance of method stability in NO and ON optimizations. EBI@CO's robustness and lower iteration requirements, as demonstrated via varied initial guesses, establish it as a highly reliable and efficient choice for achieving accurate and consistent convergence.

\subsection{Comprehensive Testing on 70 Molecular Species}

\begin{figure}[!ht]
\includegraphics[width=0.5\linewidth]{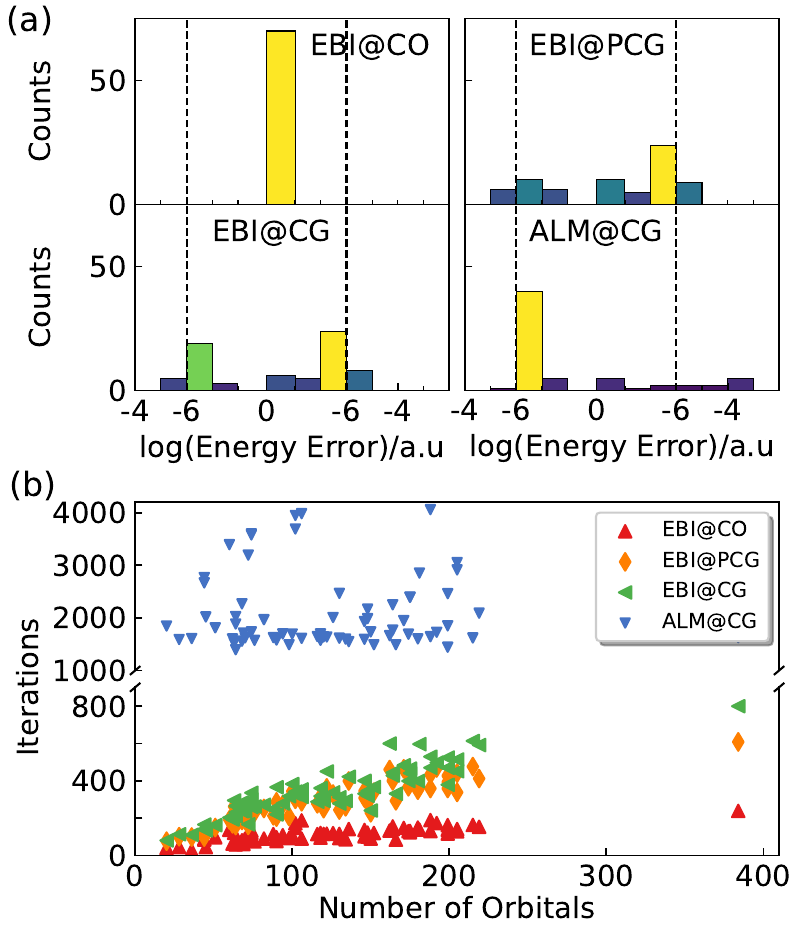}
\caption{Comparison of iterations and energy errors for EBI@CO, EBI@PCG, EBI@CG, and ALM@CG across a set of 70 molecules utilizing the $\omega$P22 functional. Panel (a) depicts the distribution of energy errors, with color intensity representing frequency—lighter shades indicate higher occurrence rates. The x-axis scales logarithmically with the magnitude of energy errors, where values to the left of zero represent negative errors, and those to the right, positive errors. Panel (b) illustrates the relationship between the total number of iterations required for convergence and the count of NOs present in each molecule. Different basis sets are employed based on the system's characteristics: the aug-cc-pVDZ set for 10 molecules involving weak interactions, def2-TZVP for 10 systems containing metal elements, def2-SVP for 10 transition metal complexes, and cc-pVTZ for the remaining 40 molecules.}
\label{fig:sum}
\end{figure}

\begin{table}[!b]
\caption{Summary of iterations required and energy errors for 70 molecules analyzed using the $\omega$P22 functional, comparing EBI@CO with various decoupled methods (EBI@PCG, EBI@CG, and ALM@CG). Reference energy values are obtained from EBI@CO calculations. Energy errors are relative to these reference values, with the unit of energy reported in a.u.}
        \begin{tabular}{ccccc}
            \hline
            \hline
        &    EBI@CO   &   EBI@PCG     &   EBI@CG      &  ALM@CG    \\
        \hline 
        \hline 
        Steps  &    113.97 	&   291.80   	&   340.96   	&  2067.04   \\
        Error  &     -    	&   8.61$\times10^{-8}$	&   2.90$\times10^{-8}$	&  3.66$\times10^{-4}$  \\
        \hline 
        \hline 
        \end{tabular} \label{tab:sum}
\end{table}

Building upon previous findings demonstrating the enhanced convergence speed and stability of EBI@CO across various functionals, basis sets, and initial conditions, we extend our investigation to its performance across a diverse set of 70 molecules using the $\omega$P22 functional \cite{Ai2022jpcl,LYPr}. Fig. \ref{fig:sum} highlights the distribution of energy errors and iteration counts, revealing that for the majority of molecules, the energy deviations between EBI@CO and decoupled methods remain within $10^{-6}$, with EBI@CO typically achieving lower energy levels. Interestingly, in the case of free radical systems, decoupled methods employing EBI occasionally register slightly lower energies than EBI@CO. However, the marginal deviations, approximately $10^{-6}$, underscore the competitive performance of EBI@CO even in these challenging scenarios. ALM@CG encounters notable difficulties in converging for free radical systems, often requiring 3000-4000 iterations to reach convergence, with the results frequently corresponding to local minima rather than global ones. Table \ref{tab:sum} summarizes the average iterations and energy errors for all analyzed systems, contrasting the efficiencies of different optimization methods. EBI@CO stands out significantly, converging within an average of 113.97 iterations to a precision of $10^{-8}$, showcasing a dramatic 95\% reduction in computational effort compared to the 2067.04 iterations required on average by decoupled methods. This comprehensive analysis not only affirms the superior efficiency and stability of EBI@CO across a wide array of molecular systems but also highlights its potential to significantly reduce computational costs in RDMFT calculations.

\subsection{Optimization Challenges for Large Molecular Systems}

\begin{figure}
\includegraphics[width=0.7\linewidth]{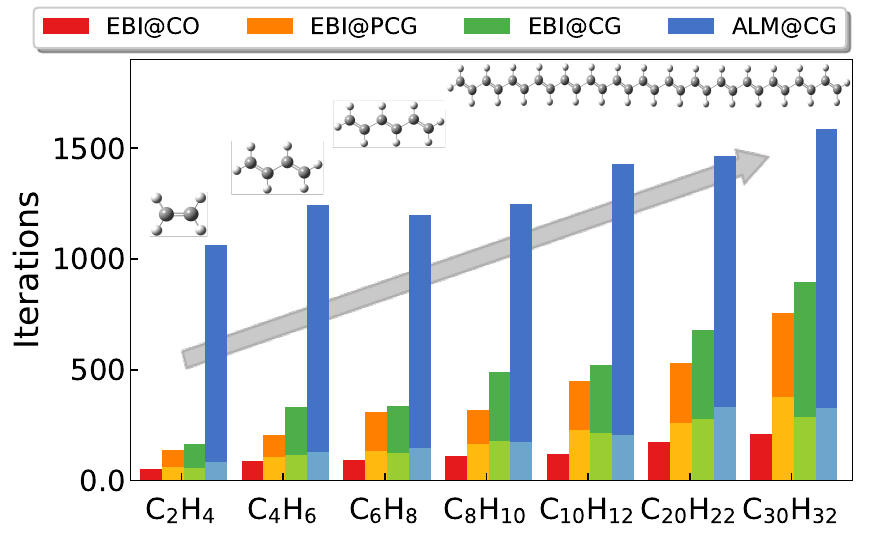}
\caption{Comparison of the number of iterations of EBI@CO and decoupled methods for olefins from $\mathrm{C_2H_4}$ to $\mathrm{C_{30}H_{32}}$. The light and dark colors represent the iterations of NO and ON optimizations, respectively. The calculations are performed with the $\omega$P22 functional and the cc-pVDZ basis set.}
\label{fig:alkene-wp22}
\end{figure}
    
\begin{figure}
\includegraphics[width=0.5\linewidth]{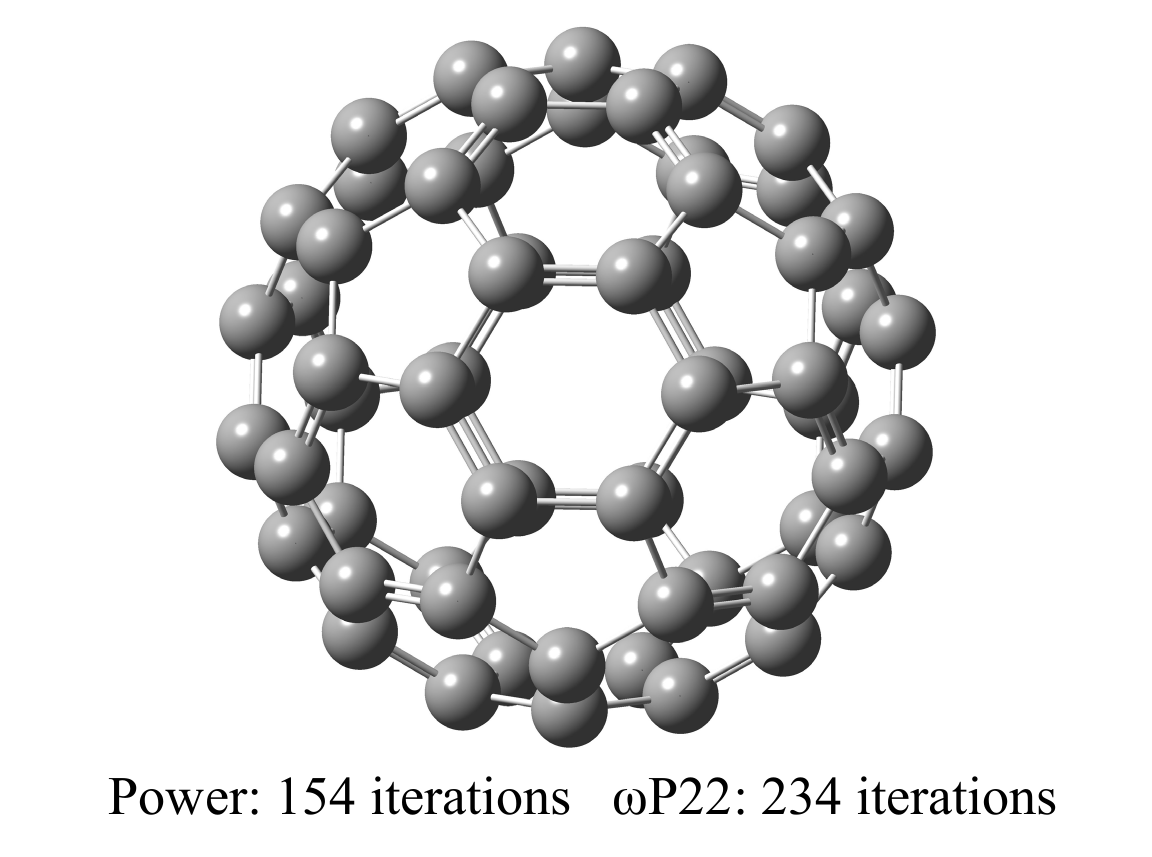}
\caption{The structure of $\mathrm{C_{60}}$ and the iterations of EBI@CO for two functionals: power functional with $m=0.7$ and $\omega$P22 functional. The convergence threshold is $10^{-8}$.}
\label{fig:C60}
\end{figure}

The efficacy of EBI@CO in handling large-scale systems is further validated through computational experiments on olefin chains ranging from $\mathrm{C_2H_4}$ to $\mathrm{C_{30}H_{32}}$, marking a critical step towards the broader application of RDMFT to sizable molecular systems. These calculations, performed using the $\omega$P22 functional, serve as a benchmark for assessing scalability. As depicted in Fig. \ref{fig:alkene-wp22}, EBI@CO consistently necessitates fewer iterations to achieve convergence compared to the fastest decoupled method, demonstrating a minimal increase in iteration count as system size escalates. Remarkably, for the $\mathrm{C_{30}H_{32}}$ olefin, convergence to an energy precision of $10^{-8}$ is attained within merely 209 steps, showcasing the method's efficiency and scalability. The performance of EBI@CO extends to the fullerene $\mathrm{C_{60}}$, tested with both the power functional ($m=0.7$) and the $\omega$P22 functional. Results, illustrated in Fig. \ref{fig:C60}, indicate that energy convergence to $10^{-8}$ is achieved in 154 and 234 iterations, respectively. This substantiates EBI@CO's significant enhancement to the computational prowess of RDMFT, affirming its readiness for tackling large molecular systems.

\section{Concluding Remarks \label{conc}}

In this study, we introduce EBI@CO, a novel coupled optimization method that synergistically integrates the unitary optimization with the EBI method to address the convergence challenges inherent in RDMFT. A comprehensive exposition of the method's formulaic foundations and algorithmic processes is provided. Through extensive testing across a diverse array of molecules, varying initial guesses, multiple basis sets, and different functionals, EBI@CO's superior performance is unequivocally demonstrated. It surpasses existing decoupled optimization methods in terms of convergence speed, accuracy, and reliability.

Remarkably, EBI@CO achieves convergence within 154 iterations to a precision of $10^{-8}$ a.u for the complex system of the $\mathrm{C_{60}}$ fullerene, underscoring its efficacy and scalability. This work enhances the computational feasibility of RDMFT and paves the way for the practical application of the recently developed HyperComplex Kohn-Sham (HCKS) theory \cite{HCKS,Su2022es,Zhang2023pra}. The advancements presented herein not only bolster RDMFT's position as a viable tool for quantum chemical analysis but also broaden its applicability and scope for future research and development.

\begin{itemize}
\item Supporting Information.  Algorithms of decoupled optimization methods and more results of different optimization methods 
\end{itemize}

\begin{acknowledgement} 
Support from the National Natural Science Foundation of China (Grants No. 22122303 and No. 22073049) and Fundamental Research Funds for the Central Universities (Nankai University: No. 63206008) is appreciated.
\end{acknowledgement}

\bibliography{refnew}

\end{document}